\documentclass{nature}

\usepackage{lineno}
\usepackage{lmodern}
\usepackage{amsmath}
\usepackage{amssymb}
\usepackage{graphicx}
\usepackage{hyperref}
\usepackage{xspace}
\usepackage{caption}
\usepackage{mathptmx}
\usepackage{mathrsfs}
\usepackage{pdfpages}
\usepackage{mathtools} 
\usepackage{bm}
\usepackage{upgreek}

\usepackage{xr-hyper}
\makeatletter
\newcommand*{\addFileDependency}[1]{
  \typeout{(#1)}
  \@addtofilelist{#1}
  \IfFileExists{#1}{}{\typeout{No file #1.}}
}
\makeatother

\newcommand*{\myexternaldocument}[1]{
    \externaldocument{#1}
    \addFileDependency{#1.tex}
    \addFileDependency{#1.aux}
}

\myexternaldocument{supp_info}

\title{Ultrafast laser-driven topological spin textures on a 2D magnet
}

\author{Mara Strungaru$^1$, Mathias Augustin$^{2}$, Elton J. G. Santos $^{1,3\dagger}$}

\makeatletter
\let\saved@includegraphics\includegraphics
\AtBeginDocument{\let\includegraphics\saved@includegraphics}
\renewenvironment*{figure}{\@float{figure}}{\end@float}
\makeatother

\begin{document}

\maketitle

\begin{affiliations}
\item Institute for Condensed Matter Physics and Complex Systems, School of Physics and Astronomy, The University of Edinburgh, EH9 3FD, United Kingdom 
\item Donostia International Physics Center (DIPC), 20018 Donostia-San Sebasti\'an, Basque Country, Spain
\item Higgs Centre for Theoretical Physics, The University of Edinburgh,  EH9 3FD,  United Kingdom \\
$^{\dagger}$Corresponding author: esantos@ed.ac.uk
\end{affiliations}

\date{}

\begin{abstract}
Ultrafast laser excitations provide an efficient and low-power consumption alternative since different magnetic properties and topological spin states can be triggered and manipulated at the femtosecond (fs) regime.  However, it is largely unknown whether laser excitations already used in data information platforms can manipulate the magnetic properties of recently discovered two-dimensional (2D) van der Waals (vdW) materials. Here we show that ultrashort laser pulses (30$-$85 fs) can not only manipulate magnetic domains of 2D-XY CrCl$_3$ ferromagnets, but also induce the formation and control of topological nontrivial meron and antimeron spin textures. We observed that these spin quasiparticles are created within $\sim$100 ps after the excitation displaying rich dynamics through motion, collision and annihilation with emission of spin waves throughout the surface. Our findings highlight substantial opportunities of using photonic driving forces for the exploration of spin textures on 2D magnetic materials towards magneto-optical topological applications.

\end{abstract}


\section*{Introduction}

The discovery of graphene has pioneered the study and development of technological applications based on layered materials with appealing properties such as high flexibility and optical transparency\cite{Novoselov12}. Recently, magnetism has been discovered at the monolayer regime which opened a rapid exploration of a wide class of compounds containing ferromagnets\cite{Gong2017, huang2017layer}, antiferromagnets\cite{Lee2016,Lee2016_1,Alliati20} towards disparate design platforms\cite{Genome22}. Multiple layers of 2D materials can show intriguing properties such as crystal and layer dependent magnetic properties\cite{huang2017layer,GuguchiaSantos} or unusual magnetic ground-states via twist engineering\cite{Xie2022,Genome22}. Since magnetic materials\cite{huang2017layer,Gong2017,GuguchiaSciAdv,wahab2021quantum,GuguchiaSantos,Wahab21,CantosPrieto2021,Alliati20,Genome22,Jenkins22} are highly used in technological applications, a natural question arises: what is the most efficient way to control the magnetism in such systems. Recent studies of  current-driven domain wall dynamics\cite{Alliati20,Wahab21} or gate-controlled anisotropy\cite{verzhbitskiy_controlling_2020} can lead to low-power consumption, and high-speed devices. 

Other alternatives through laser approaches\cite{Carbone18,Ezawa13,Raising07,Beaurepaire09,Koopmans2010} also provide energy-efficient means to manipulate the magnetic properties via demagnetization, spin-reorientation, or even modification of magnetic structures at a short timescale\cite{Raising07,KIMEL2020}. Initially applied on elemental magnets\cite{beaurepaire1996}, ultrafast laser pulses have led to crucial discoveries including spin switching\cite{Rasing05,ostler2012ultrafast,Dabrowski:2022aa}, all-optical reversal\cite{Rasing07} and manipulation of topological properties on magnetic nanostructures\cite{Finazzi2013,Carbone18}. Despite these achievements, 2D vdW magnets are largely unexplored through ultrafast laser excitations, in particular the behavior of fundamental quantities such as magnetisation, magnetic domains and how to induce the appearance of strongly correlated phenomena remain yet to be elucidated. Here we use CrCl$_3$ vdW magnet as a sample system and demonstrate the formation and control of topologically non-trivial vortex quasiparticles through laser radiation in the time-scale used in lab measurements. CrCl${_3}$ is a popular 2D magnetic material\cite{Mandar21,Xiadong19,Wee20} with XY ferromagnetic order (Fig.\ref{fig1}{\bf a}) at the monolayer limit which can host merons and antimerons intrinsically in its magnetic structure\cite{Augustin2021}. The delicate interplay between strong in-plane dipole-dipole interactions and the weak out-of-plane magnetic anisotropy allows such quasiparticles to appear during zero-field cooling. 
Despite of this phenomenon, we show that 
the effective external torque given by the ultrafast laser excitation provided enough heat to overcome the exchange energy and reverse the spin orientations into specific topologies leading to a robust driving-force for generating vortices and anti-vortices in CrCl$_3$.


\section*{Results} 

\subsection{A biquadratic spin Hamiltonian and the two-temperature model.}    
    
We model the system through atomistic spin dynamic simulations\cite{wahab2021quantum,kartsev2020biquadratic} with interactions being described by an all-round spin Hamiltonian: 
\begin{equation}
\mathcal{H} = -\frac{1}{2}\,  \sum_{i,j }  \mathbf{S}_i \mathcal{J}_{ij} \mathbf{S}_j  -\frac{1}{2}\,  \sum_{i,j } K_{ij}\, (\mathbf{S}_i\, \cdot  \mathbf{S}_j\:)^2- \sum_{i} D_i (\mathbf{S}_i\, \cdot  \mathbf{e})^2 - \sum_{i} \mu_i  \mathbf{S}_i\, \cdot  \mathbf{B}_{\rm dp} 
\label{gen_ham}
\end{equation}
where $i$, $j$ represent the atoms index, $\mathcal{J}_{ij}$ represents the exchange tensor that for CrCl${_3}$ contains only the diagonal exchange terms,  
$K_{ij}$ is the biquadratic exchange interaction\cite{kartsev2020biquadratic}, $D_i$ the uniaxial anisotropy, which is orientated out of plane ($\mathbf{e}=(0,0,1)$) and $\mathbf{B}_{\rm dp}$ is the dipolar field calculated via the macrocell method\cite{wahab2021quantum} using a cell size of $2$ nm. The exchange interactions for CrCl${_3}$ have been previously parameterized from first-principles calculations\cite{kartsev2020biquadratic,Augustin2021} and contains up to three nearest neighbors. The inclusion of biquadratic exchange \cite{kartsev2020biquadratic} and next-nearest exchange interactions leads to the stabilisation of non-trivial spin structures\cite{Augustin2021,Luis21} as previously demonstrated. 

{The magnetisation dynamics is obtained by solving the Landau-Lifshitz-Gilbert (LLG) equation applied at the atomistic level:
\begin{equation}
 \frac{\partial \mathbf{S}_i }{\partial t}   = -\frac{\gamma}{(1+\lambda^2) } \mathbf{S}_i \times  (\mathbf{H}_{i}+   \lambda \mathbf{S}_i \times \mathbf{H}_{i} )   
 \label{llg}
\end{equation}
where $\lambda$  represents the coupling to the heat bath, $\gamma$ the gyromagnetic ratio and $\mathbf{H}_{i}$ the effective field that acts on the spin $\mathbf{S}_i$ . The effective field can be calculated from the Hamiltonian of the model to which we add a thermal noise $ \xi_i$:
\begin{equation}
    \mathbf{H}_i= - \frac{1}{\mu_i \mu_0} \frac{\partial {\mathcal{H}} }{\partial \mathbf{S}_i} + \xi_i
\label{LLG2} 
\end{equation}
The thermal field is assumed to be a white noise, with the following mean, variance and strength, as calculated from the Fokker-Planck equation: 
\begin{equation}
 \langle \xi_{i\alpha}(t) \rangle=0, ~~  \langle \xi_{i\alpha}(t)\xi_{j\beta}(s) \rangle= 2D \delta_{\alpha,\beta} \delta_{ij} \delta(t-s)
\label{eff}
\end{equation}
\begin{equation}
D= \frac{ \lambda  k_{\rm B} T }{\gamma \mu_{\rm i} \mu_{\rm 0}}
\end{equation}
where $T$ represents the thermostat temperature, $\mu_{\rm i}$ the magnetic moment and $\mu_{\rm 0}$ the magnetic permeability.}

We include the effect of the laser pulse via the two-temperature model (2TM)\cite{chen2006semiclassical} which couples the electronic and phonon bath via: 
\begin{equation}
C_{\rm e0}T_{\rm e} \frac{dT_{\rm e}}{dt}=-G_{\rm ep}(T_{\rm e}-T_{\rm p})+P(t)
\end{equation}
\begin{equation}
C_{\rm p}\frac{dT_{\rm p}}{dt}=-G_{\rm ep}(T_{\rm p}-T_{\rm e}) - \kappa_{\rm e} \nabla T_{\rm p}
\label{eq_sink}
\end{equation}
where $C_{\rm e0}$ and $C_{\rm p}$ are the electron and phonon heat capacity, respectively;   
$G_{\rm ep}$ represents the electron-phonon coupling factor; 
$T_{\rm p}$, $T_{\rm e}$ are the phonon and electron temperatures, respectively; 
$\kappa_{\rm e}$ is the diffusion coefficient and $P(t)$ is the time dependent laser pulse power. 

The electronic temperature is coupled to the magnetic system through the thermal field entering into the Landau-Lifshitz-Gilbert (LLG) equation. The laser power density takes a Gaussian form:
\begin{equation}
    P(t)=\frac {2F_{\rm 0}}{\delta t_{\rm p}\sqrt{\pi/\ln2}} \exp[(-4\ln2)(\frac{t}{t_{\rm p}})^2]
\end{equation}
where $F_{\rm 0}$ is the laser fluence (in units of energy density), $t_{\rm p}$ is the pulse temporal width and $\delta$ is the optical penetration depth, assumed to be $\delta =$ 10 nm. The 2TM can be extended to include a term to reflect the heat diffusion to the substrate, via a heat-sink coupling term of 1/(100 ps). The heat diffusion to the substrate has a time-scale in the order of pico- to nano-seconds and in the model is included as $-\kappa_{\rm e} \nabla T_{\rm p}$ in Eq.\ref{eq_sink}. {The parametrization of the 2TM (Supplementary Table 1) 
follows that used in the description of the experimental magnetisation dynamics recently measured for a parent halide compound\cite{padmanabhan2020coherent} 
and modelled via a three-temperature approach ($T_{\rm p}$, $T_{\rm e}$, $T_{\rm s}$). 
Since the spin dynamics is included directly into our model, we need to consider only two temperatures ($T_{\rm p}$, $T_{\rm e}$) instead. The evolution of the spin temperature ($T_{\rm s}$) will be given directly by the magnetisation dynamics. Parameters such as $C_{\rm e0}$, $C_{\rm p}$ and $G_{\rm ep}$ are considered temperature independent in the 2TM and can be extracted from the thermal conductivity, electronic specific heat, and phonon specific heat\cite{McGuire2015,Boeckl20}.  
The electron-phonon coupling is approximated as $C_{\rm e}T_{\rm e}/t_{\rm e-ph}$, where  $t_{\rm e-ph}$ is the electron-phonon thermalisation time. Since CrCl$_3$ has a low Curie Temperature ($T_{\rm C}=19$ K\cite{Augustin2021}), we will approximate the electron-phonon coupling as that at $T=10$ K. The values of thermal bath coupling ($\alpha=0.1$) and heat sink coupling ($\tau=100 $~ps) shown in Supplementary Table 1 have been chosen to allow fast numerical simulations. The pulse duration of $t_{\rm p}=100 $~fs is a typical laser pulse width\cite{ostler2012ultrafast}, however this value is highly dependent on experimental capacities. 
}


\subsection{Ultrafast spin dynamics on CrCl$_3$.} 

We observe that as the system is excited with a short laser pulser (85 fs) with an energy fluence of 0.01 mJ cm$^{-2}$ the initial in-plane magnetisation $M_{\rm t}=\sqrt{M_{\rm x}^2+M_{\rm y}^2}$ reduces rapidly from its saturated state ($M_{\rm t}/M_{\rm s}=1$) to a minimum near zero within 25 ps (Fig. \ref{fig1}{\bf b}). The demagnetisation process is noticed to be barely dependent on the applied fluence (inset in Fig. \ref{fig1}{\bf b}) showing a similar demagnetised state behaviour. A close look at the variation of the temperatures with time indicates that the system peaks at T$_{\rm e}=60$ K during the laser pulse (Fig. \ref{fig1}{\bf c}) which is larger than the Curie temperature $T_{\rm C}=19$ K\cite{Augustin2021} of the monolayer CrCl$_3$. Since the vdW layer is coupled to a heat sink, the energy deposited is quickly dissipated through the substrate, reaching T$_{\rm e}=$T$_{\rm p}\sim$0.10 K after 200 ps. During this thermal relaxation process, an increase in both transversal M$_{\rm t}$ and out-of-plane M$_{\rm z}$ magnetisation components is observed due to the decreased temperature. However, the saturated in-plane magnetisation is not recovered after the laser pulse, but rather it breaks into magnetic domains (Fig. \ref{fig2}{\bf a-e}) leading to a transversal magnetisation of M$_{\rm t}/$M$_{\rm s}=$0.2 at longer times. We noticed that after the application of the laser pulse, a small out-of-plane magnetisation M$_{\rm z}/$M$_{\rm s}=$0.04 is developed. By looking at the spin maps (Fig. \ref{fig2}{\bf f-j}) we identify small circular areas distributed randomly around the surface where the non-zero magnitudes of M$_{\rm z}/$M$_{\rm s}$ are centered after 200 ps. The polarisation of those areas is an indication of the creation of merons and antimerons quasiparticles on the surface. 



\subsection{Topological number as a descriptor.}

We can further characterise those spins structures created by the laser pulse through the topological number\cite{Augustin2021,eriksson_1990,rozsa_prb}:   
\begin{equation}\label{topological-N}
N = \frac{1}{4\pi}\int  \mathbf{{n}} \cdot \left(\frac{\partial\mathbf{{n}}}{\partial x} \times\frac{\partial \mathbf{{n}}}{\partial y}  \right) d x ~d y 
\end{equation}
where ${\mathbf n}$ is the direction vector of magnetisation ${\bf M}$, e.g., ${\mathbf n}=\frac{\bf M}{|\bf M|}$.   
Eq.~\ref{topological-N} can also be represented by\cite{Fisher04} $N=wp/2$ indicating a product between the 
vorticity or winding number $w= \pm1$ (which determines the in-plane components of the magnetic 
moment) with the polarity $p=\pm1$ (which defines the out-of-plane core polarisation). 
For different combinations of $w$ and $p$, standard spin configurations are those for merons ($N=-1/2$) and antimerons ($N=1/2$). However, higher order quasiparticles with $w=\pm2$ are possible which resulted in second-order antimerons ($N=1$) and second-order merons ($N=-1$)\cite{Krol:21}. 
By looking at the local profile of the magnetisation after the laser excitation (Fig. \ref{fig3}{\bf a-d}) 
we can identify several non-trivial topological spin textures characteristic of meron and antimerons.
The slightly smaller values computed for the spin features observed within the range of 
$N\sim 0.40-0.45$ is due to the finite area considered on the integration of Eq.~\ref{topological-N}.\cite{eriksson_1990,Augustin2021}. 
However, the clear spin distributions with the magnetic moment at the core 
pointing out-of-plane and the surrounding perimeter with spins aligning in-plane hallmarked the formation of merons and antimerons.   
More complex quasiparticles composed by two antimerons ($N=0.95$) are also possible to be generated during the thermal 
equilibration (Fig. \ref{fig3}{\bf d}). 

It is worth mentioning that the formation of these spin textures in CrCl$_3$ is not related with the presence of non-collinear Dzyaloshinskii-Moriya interactions not considered in the spin Hamiltonian in Eq.~\ref{gen_ham}, but on the interplay between the laser pulse heat and the system thermal equilibration. As the laser pulse quickly quenches the total magnetisation, magnon localisation takes place at small areas of the surface due to short-range exchange interactions\cite{Nowak08,Barker13}. This makes the magnetisation increases with the nucleation of magnetic droplet solitons with unstable spin textures\cite{Hoefer14,Grychtol16} perpendicular and/or parallel to the easy-plane of CrCl$_3$. Within a time scale of a few picoseconds, these droplets can split, merge and scatter until thermal equilibration is reached which induced the formation of more stable merons and antimerons with a defined spin configuration at longer times ($>$ 400 ps). {We studied as well whether thermal effects present in cooling processes can help in the creation of meron and antimerons (Supplementary Figure 1). We noticed the formation of these spin textures at all cooling times considered (0.1 ns, 0.5 ns, 3 ns and 4 ns) suggesting that indeed they can also be produced via thermal effects. The cooling time however will influence the amount of merons/antimerons present in the system, with faster cooling leading to the creation of more spin textures. Hence a fast heating or cooling as the one provided by the laser pulse can lead to a higher number of merons or antimerons, in comparison to a slow cooling process.
}



We next investigate the long time-scale dynamics of the topological spin structures that are created with the ultrafast laser pulse (Fig. \ref{fig4} and Supplementary Movie S2). We can observe different collision interaction scenarios throughout the layer at different time frames and spatial locations. Such as, at the selected area 1 in Fig. \ref{fig4}{\bf a,e} which shows a surprising annihilation dynamics followed by the emission of a spin-wave isotropically along the surface shortly after the collision (Supplementary Movie S2). We  noticed events involving pairs of multiple vortex and antivortex (area 2, Fig. \ref{fig4}{\bf b,f}) with the subsequent annihilation and spin-wave emission happening at each pair separately (areas 2-3, Fig. \ref{fig4}{\bf b,c,f,g}) and at different time frames (1.5 ns, 1.75 ns). This suggests a 1:1 vortex-antivortex relation for the emission of spin-waves despite the number of quasiparticles involved. Simultaneous collisions can also occur as displayed in area 4 (Fig. \ref{fig4}{\bf c,h}) with no apparent correlations on other events. The relative distance between the vortex and antivortex plays a role on the lifetime of the spin textures. We observe that the vortex/antivortex pair in area 5 (Fig. \ref{fig4}{\bf d,i}) survived longer relative to other events ($>$2 ns) and has a more complex dynamics given by a precession
motion. Since such spiraling orbit is present during the annihilation process for timescales beyond the simulation time ($\sim$20 ns), there is a probability that the vortex/antivortex pair might be annihilated.



Additionally we can track down the variation of the different topological spin textures via the variation of $N$ (Fig. \ref{fig4}$-$\ref{fig5}) as a function of time. The large fluctuations of $N$ observed at early stages of the equilibration ($\sim$5$-$120 ps) are due to the increased thermal fluctuations after the laser pulse (Fig. \ref{fig5}{\bf a}). At later times ($>$300 ps) $N$ varies in a much smaller scale (Fig. \ref{fig5}{\bf b}) which allows to identify unit steps correlating directly with the presence of vortex-antivortex pairs (Fig. \ref{fig5}{\bf c-f}). The annihilation phenomena appear at specific moments during the dynamics at a time-scale even beyond 1 ns. Hence the process of creation and annihilation of magnetic merons or antimerons can be tracked down via the temporal variation of $N$ which in principle is general for any spin textures on 2D magnets. A recent protocol\cite{Hesjedal17} involving X-ray scattering techniques could be adapted to the measurement of $N$ via the scattering signal from the spin textures.


\section*{Discussion}

The discovery of laser-induced topologically non-trivial merons and antimerons quasiparticles on 2D CrCl$_3$ magnets creates open routes for tailoring the magnetic properties of other vdW materials holding similar features. As more materials have been isolated and implemented on device platforms\cite{Genome22}, our findings indicate that other layered compounds may develop laser-driven topological spin textures following the recipes included here. The laser pulse heating is clearly the main driving force for the formation of the quasiparticles which occurs within the experimental regime of observation. On the particular case of CrCl$_3$ with a low Curie temperature (19.4 K)\cite{kartsev2020biquadratic}, it is necessary to couple the system to a heat-sink so the energy deposited by the laser pulse is dissipated via a substrate. Such approach induces an efficient stabilisation of vortices and antivortices on CrCl$_3$ during the equilibration process. This suggests that in order to observe topological spin textures on atomically thin vdW magnets the appropriate substrate needs to be selected and implemented on the stacking. A few options in terms of insulating substrates, i.e. BN\cite{Luhua19}, which hold high thermal conductivity may provide solution to avoid strong interactions with underneath atoms. As a matter of fact, a recent molecular beam epitaxy (MBE) of CrCl$_3$ on graphene/6H-SiC(0001)\cite{bedoya} has been fabricated which provides chemical, mechanical stability to measure the magnetic features via pumb-probe techniques. These include the utilisation of an ultrafast laser setup able to induce rapid heating of the CrCl$_3$ 
with a consequent modification of the magnetisation which can be measured through magneto-optical Kerr effect\cite{KIMEL2020}. In principle such approach involving good quality MBE CrCl$_3$ samples and a high-energy femtosecond laser would be able to probe the topological features of the system and unveil the presence of merons and antimerons on a 2D XY vdW magnet.

\section*{Methods}
We model the system through atomistic spin dynamic simulation methods \cite{wahab2021quantum,kartsev2020biquadratic} with interactions  described by a biquadratic spin Hamiltonian (Eq.\ref{gen_ham}).  All methods are included in the main text with additional details at Supplementary Information.

\section*{Data Availability}

The data that support the findings of this study are available within the paper and upon reasonable request.  


\section*{Acknowledgement}
EJGS acknowledges computational resources through CIRRUS Tier-2 HPC 
Service (ec131 Cirrus Project) at EPCC (http://www.cirrus.ac.uk) funded 
by the University of Edinburgh and EPSRC (EP/P020267/1); 
ARCHER UK National Supercomputing Service (http://www.archer.ac.uk) {via} 
Project d429, and the UKCP consortium (Project e89) 
funded by EPSRC grant ref EP/P022561/1.   
EJGS acknowledge the Spanish Ministry of 
Science's grant program ``Europa-Excelencia'' under 
grant number EUR2020-112238, the EPSRC Early Career 
Fellowship (EP/T021578/1), and the University of 
Edinburgh for funding support. 
For the purpose of open access, the author has applied a Creative Commons Attribution (CC BY) licence to any Author Accepted Manuscript version arising from this submission. 






\subsubsection*{Author Contributions} 
EJGS conceived the idea and supervised the project. 
MS performed the atomistic simulations under the supervision of EJGS. MA contributed on the discussion, and analysis. 
EJGS wrote the paper and prepared the figures with an initial 
draft prepared by MS. 
All authors contributed to this work, read the 
manuscript, discussed the results, and agreed on 
the included contents.

\section*{Competing interests}

The Authors declare no conflict of interests.


\section*{References}


\section*{Figure captions}

\setcounter{figure}{0}

\begin{figure}[h]\centering   
  \includegraphics[width=0.6\columnwidth]{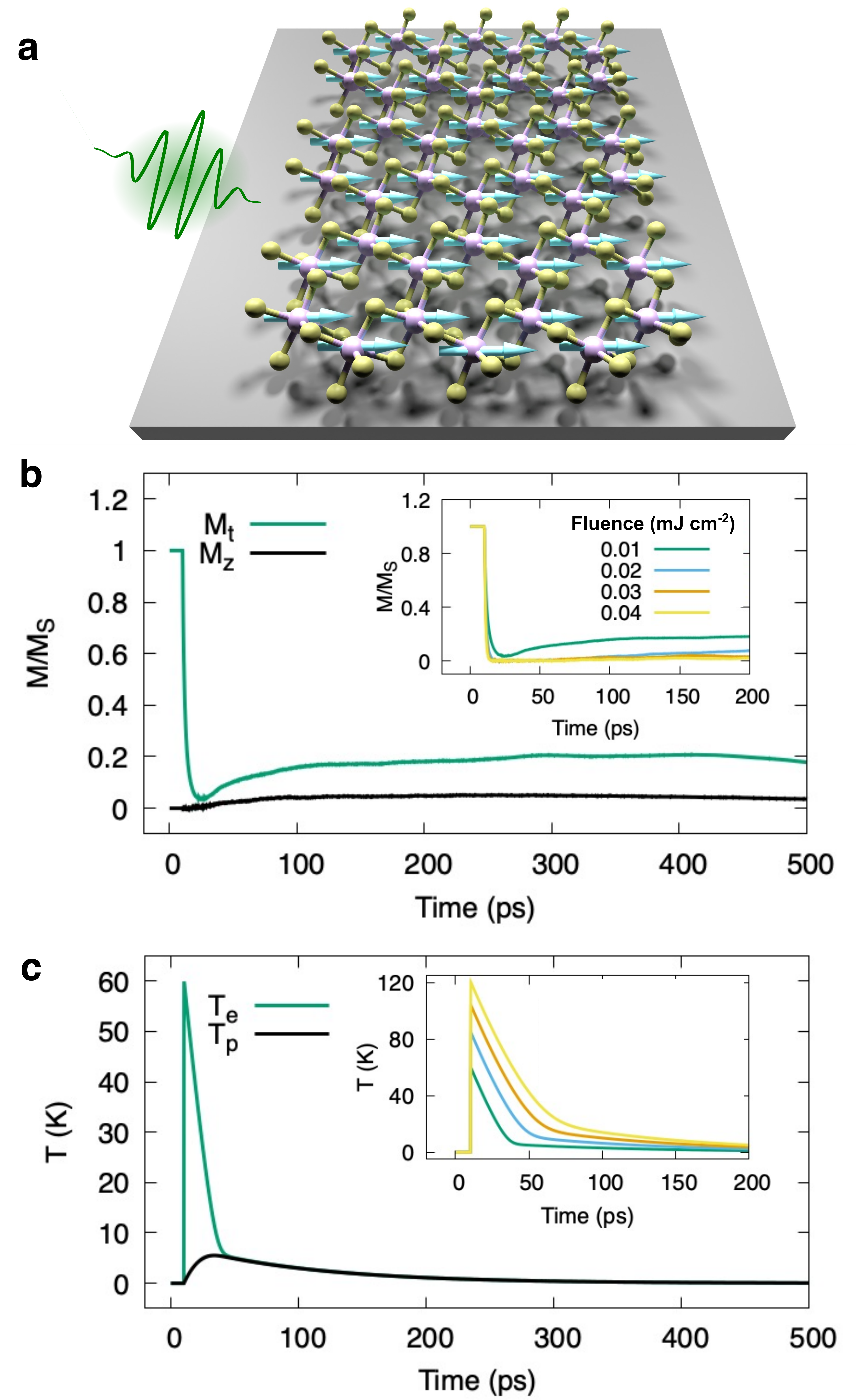}
 \caption{{\bf Excitation of 2D CrCl${_3}$ by a linearly polarised light.} 
{\bf a,} Schematic of a monolayer CrCl$_3$ with the magnetic Cr atoms represented by the pink honeycomb lattice, while the Cl atoms are given in yellow. The monolayer CrCl${_3}$ is deposited on a substrate to allow fast dissipation of the energy after the application of the laser pulse.
{\bf b,} Evolution of the in-plane $M_t=\sqrt{M_{\rm x}^2+M_{\rm y}^2}$ and out-of-plane $M_{\rm z}$ components of the magnetisation as a function of time (ps) after the  application of the laser pulse of fluence 0.01 mJ cm$^{-2}$. The inset shows $M_{\rm t}$ at different fluences (mJ cm$^{-2}$).  
{\bf c,} Evolution of the electronic $T_{\rm e}$ and phononic $T_{\rm p}$ temperatures after the laser pulse excitation (0.01 mJ cm$^{-2}$). Inset shows $T_{\rm e}$ at different fluences following labels in the inset on {\bf b}.\label{fig1}     
 } 
\end{figure}

\begin{figure}[!h]\centering   
  \includegraphics[width=0.99\columnwidth]{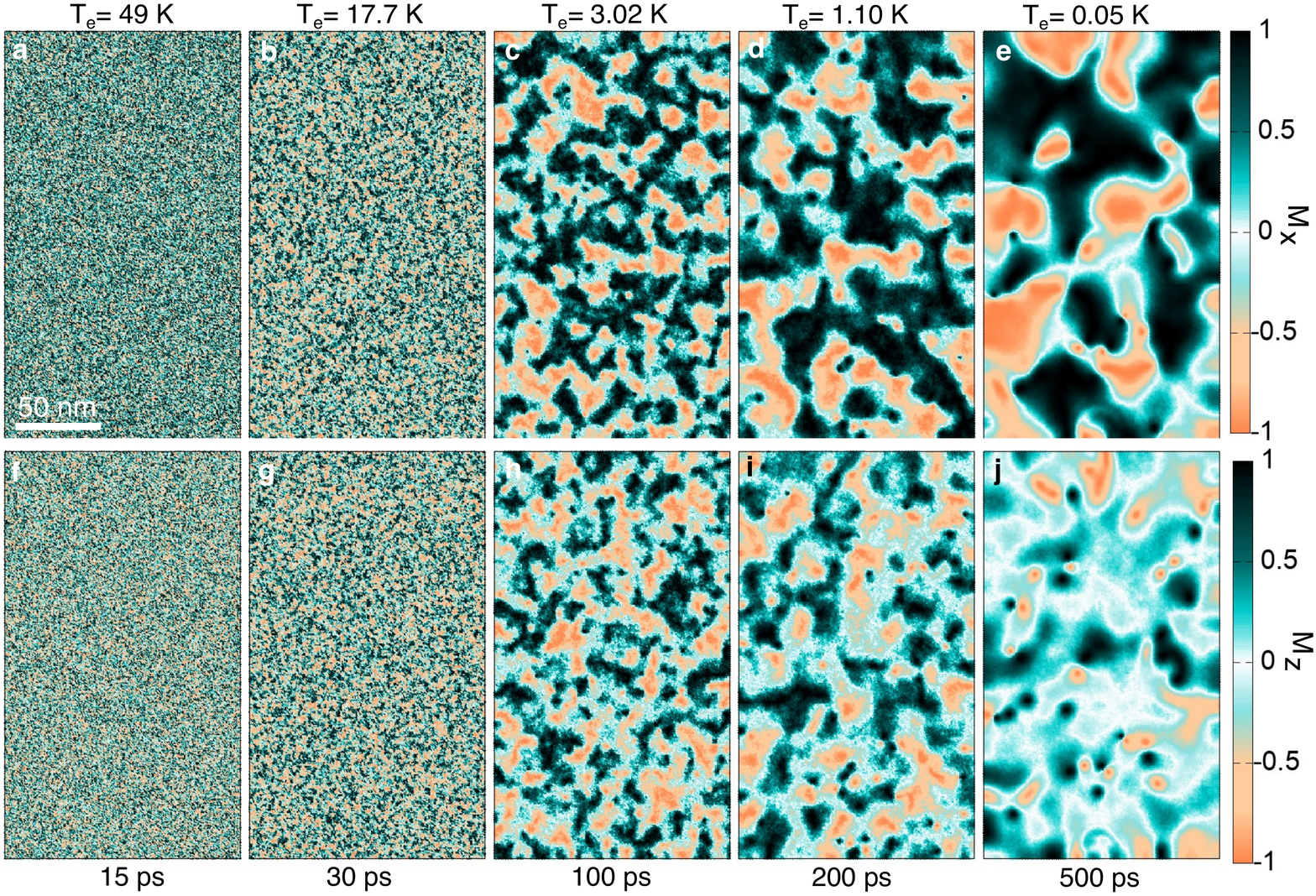}
 \caption{{\bf Magnetic evolution after laser pulse.} {\bf a-e,} Snapshots of the spin dynamics on the in-plane component M$_{\rm x}$ at different time steps (ps) and electronic temperature 
 T$_{\rm e}$ (K) of monolayer CrCl$_3$. {\bf f-j,} Similar as {\bf a-f,} for M$_{\rm z}$. 
 The small circular dark/bright spots correspond to the emergence of vortices 
 and antivortices with the laser excitation.  \label{fig2}
 } 
\end{figure}


\begin{figure}[!h]\centering   
  \includegraphics[width=1\columnwidth]{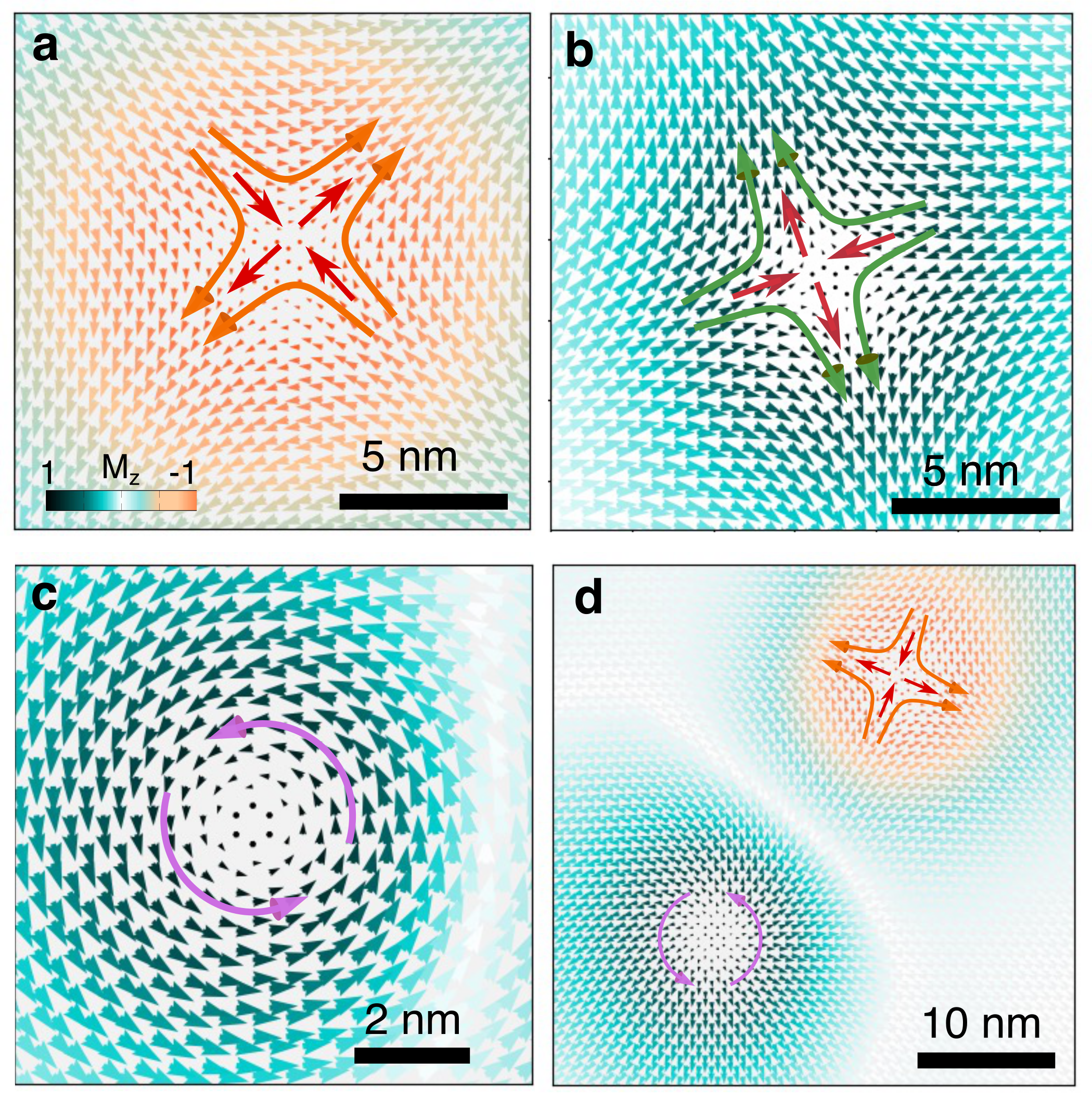}
 \caption{{\bf Vortex and antivortex creation after light excitation.} 
 {\bf a-b,} Antivortex ({vorticity} $w=-1$) formation  
 with meron ($N\sim -0.45$) and antimeron ($N\sim 0.40$) quasiparticle features, respectively. {\bf c,} Vortex ($w=+1$) stabilisation with antimeron characteristics ($N\sim 0.44$). 
 The big arrows in {\bf a-b} indicate the average behaviour of the in-plane components in the perimeter of quasiparticle.
 {\bf d,} Creation of a higher order topological spin textures ($N=0.95$) composed of two antimerons. 
 The variation of the magnitudes of $N$ is observed within $\pm$0.03 within the area considered for the integration of Eq.~\ref{topological-N}. \label{fig3}
 } 
\end{figure}

\begin{figure}[!h]\centering   
  \includegraphics[width=1\columnwidth]{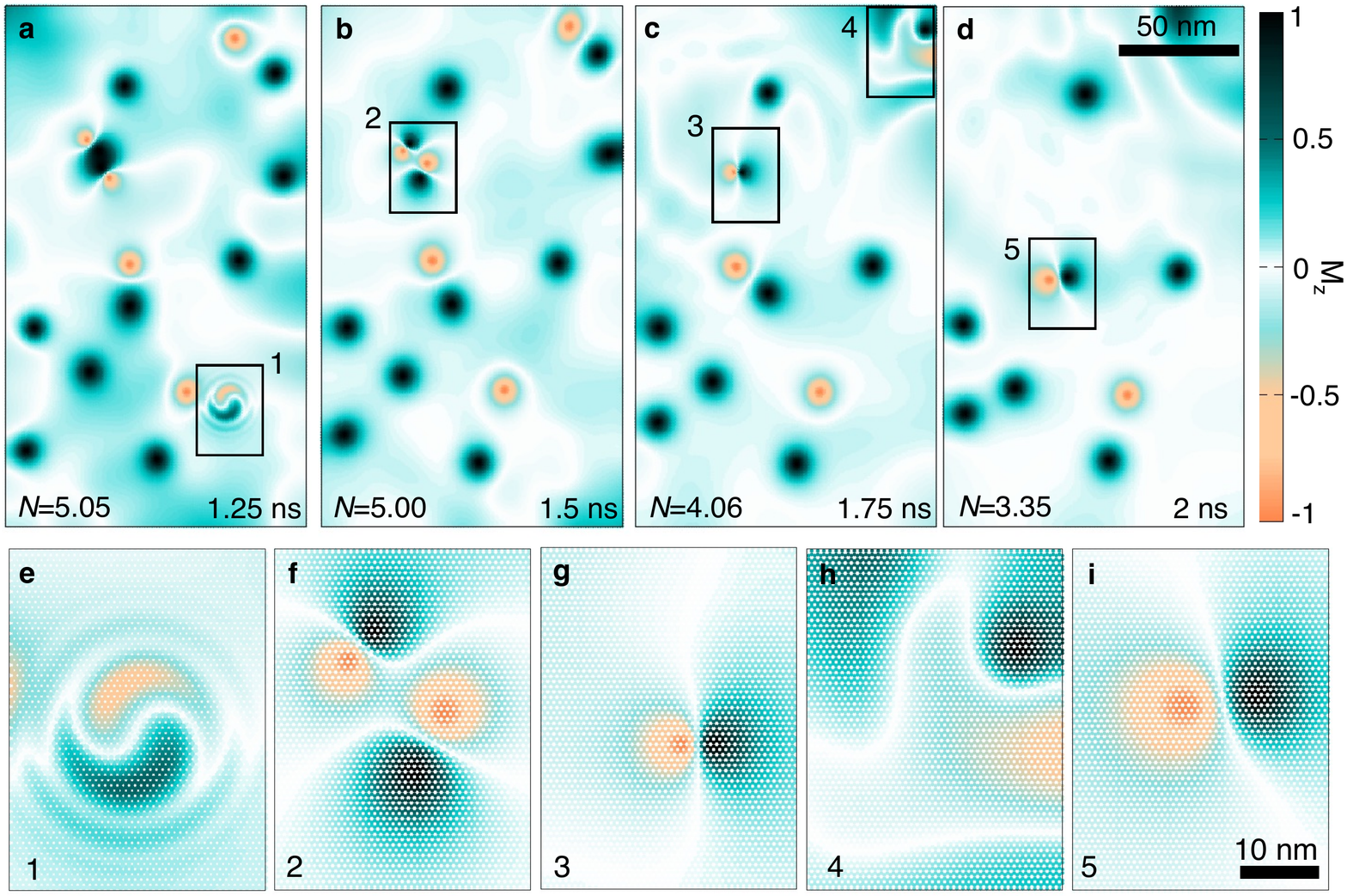}
 \caption{{\bf Large timescale evolution of merons and antimerons.} 
{\bf a-d,} Spin dynamics at long times ($1.25-2$ ns) showing the dynamics of the topological spin textures after the application of a laser pulse of fluence 0.01 mJ cm$^{-2}$. The topological number $N$ is computed at each snapshot. The small highlighted areas (1, 2, 3, 4, 5) display specific events of the interactions between vortex and antivortex. {\bf e-i,} Zoom-in at the small areas showing an annihilation event ({\bf e, h}) followed by the emission of spin-waves, collisions between two pairs of vortex and antivortex ({\bf f}), collision between single vortex and antivortex ({\bf g}) and precession of vortex and antivortex pair ({\bf i}). See Supplementary Movie S2 for additional details.  
} \label{fig4}
\end{figure}

\begin{figure}[!h]\centering   
  \includegraphics[width=1\columnwidth]{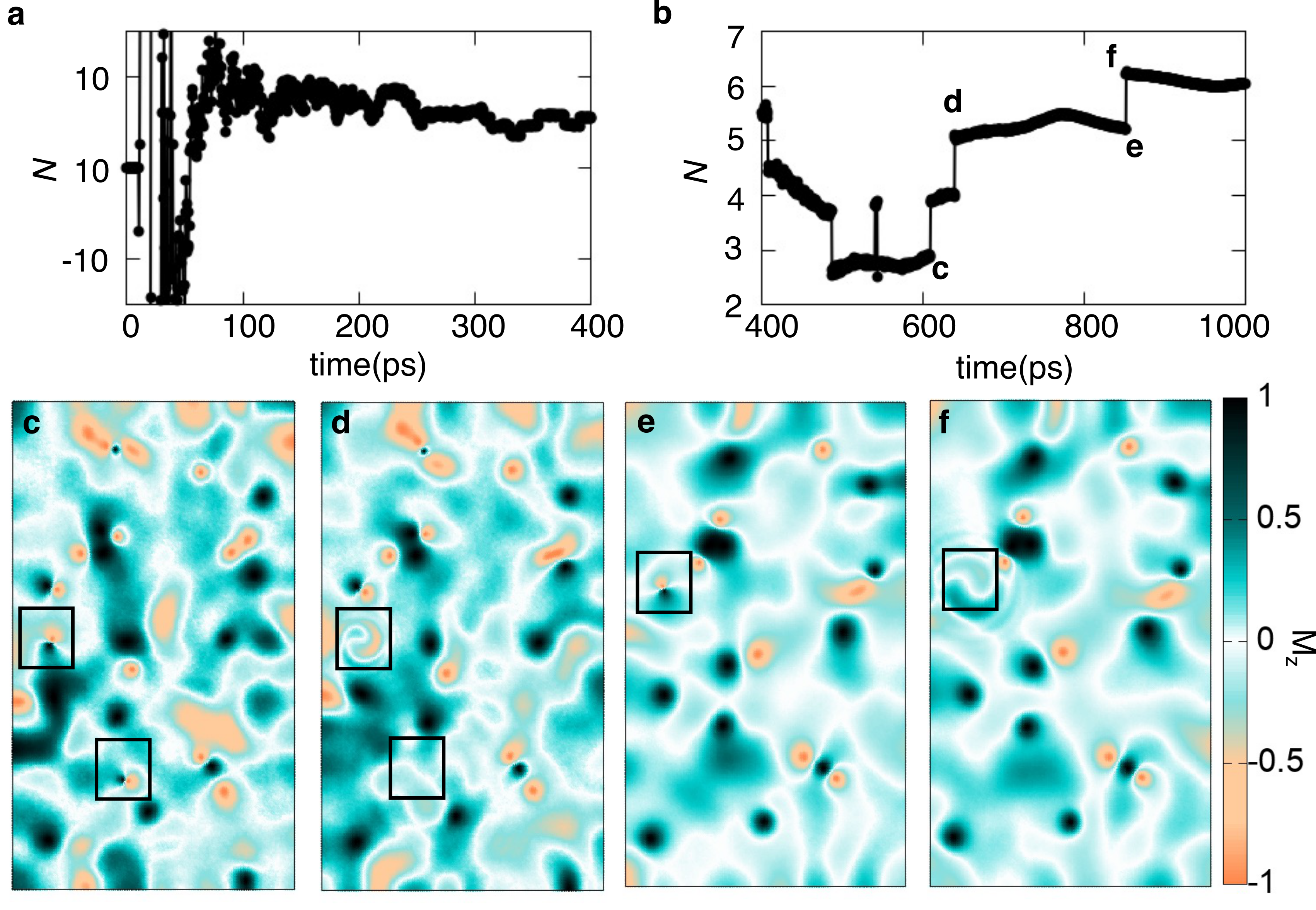}
 \caption{{\bf Topological number as a descriptor.} {\bf a-b,} Variation of the topological number 
 $N$ within 0$-$400 ps and 400$-$1000 ps intervals, respectively. Laser pulse fluence at 0.01 mJ cm$^{-2}$. 
 {\bf c-f,} Snapshots of the spin dynamics of monolayer CrCl$_3$ 
  at different times highlighted in {\bf b}. The different jumps (c, d, e, f)  
  represent selected areas where the annihilation of a vortex/antivortex pair is 
  present which is associated with an unit variation of $N$. \label{fig5}
  } 
\end{figure}

\end{document}